\documentclass[english]{revtex4}
\usepackage[T1]{fontenc}
\usepackage[latin9]{inputenc}
\usepackage{amsmath}
\usepackage{amssymb}
\usepackage{esint}

\makeatletter
\@ifundefined{textcolor}{}
{%
 \definecolor{BLACK}{gray}{0}
 \definecolor{WHITE}{gray}{1}
 \definecolor{RED}{rgb}{1,0,0}
 \definecolor{GREEN}{rgb}{0,1,0}
 \definecolor{BLUE}{rgb}{0,0,1}
 \definecolor{CYAN}{cmyk}{1,0,0,0}
 \definecolor{MAGENTA}{cmyk}{0,1,0,0}
 \definecolor{YELLOW}{cmyk}{0,0,1,0}
}

\usepackage{babel}

\makeatother

\usepackage{babel}
\begin{document}
\title{Dark Matter as gravitational solitons in the weak field limit}
\author{Torsten Asselmeyer-Maluga}
\email{torsten.asselmeyer-maluga@dlr.de}

\address{German Aerospace Center (DLR), Berlin, Germany}
\author{Jerzy Kr{\'o}l}
\email{iriking@wp.pl}

\address{University of Information Technology and Management, Chair of Cognitive
Science and Mathematical Modelling, ul. Sucharskiego 2, 35-225 Rzesz{\'o}w,Poland}
\begin{abstract}
In this paper, we will describe the idea that dark matter 
partly consists of gravitational solitons (gravisolitons).
The corresponding solution is valid for weak gravitational fields 
(weak field limit) with respect to a background metric.
The stability of this soliton is connected with the existence 
of a special foliation and amazingly with the smoothness properties
of spacetime. Gravisolitons have many properties of dark matter, such as no interaction with light but act on matter via gravitation. 
In this paper, we showed that the gravitational 
lensing effect of gravisolitons agreed with the lensing effect of usual matter. 
Furthermore, we obtained the same equation of state $w=0$ as matter. 
\end{abstract}

\maketitle

\section{Introduction}

In the 1930s, astronomers, among them the Swiss astrophysicist Fritz
Zwicky \citep{Zwicky1933,Zwicky1937}, made observations which seem
to imply that the orbits of stars in various galaxies did not have
the expected relation between speed and distance from the galactic
center. However, it took more than 30 years, until 1975, for careful studies
of several different galaxies to confirm the results of Zwicky. Using
new techniques, the astronomer Vera Rubin \citep{Rubin1980} showed
that the speeds of stars in orbit in every galaxy tended to be independent
of the distance from the center of the galaxy. However, this observation
is in sharp contrast to the theory: If the visible mass of galaxies
is responsible for star acceleration, then the speeds should vary
as $\sqrt{1/r}$, with respect to the distance. With this contradiction
in mind, astronomers and physicists quickly realized that almost all
(90\%) of the mass of galaxies resides in an invisible halo of unknown
matter, called \emph{dark matter}, sticking out into space for a distance
around 10 times the visible galactic radius. Dark matter is matter
that emits or reflects minimal to no light, but does have a gravitational
influence. Evidence for dark matter is present in 
\begin{itemize}
\item the motion of stars in galaxies, 
\item the orbits of galaxies in galaxy clusters, 
\item the temperature of intracluster gas in galaxy clusters and 
\item the gravitational lensing of distant galaxies. 
\end{itemize}
The appearance of large-scale structure (e.g., the distribution of
galaxies) is very hard to understand without Dark matter. This large
scale structure is particularly encoded into the microwave background
as measured by the satellites like Cosmic Background Explorer (COBE), Wilkinson Microwave Anisotropy Probe (WMAP) 
 or PLANCK \citep{WMAP-7-years,PlanckCosmoParameters2013}.
One way to accommodate this is to go to a dark matter model in which
you have cold dark matter (CDM) to act as a seed for galaxy formation. But
there are also results which contrast with the CDM hypothesis
like the distribution of satellite galaxies around the Milky Way \citep{MilkaywayDwarfGalaxies2009},
the absence of dark matter in our sun system \citep{DarkMatterSunsystem2012}
and the existence of a vast polar structure of satellite galaxies
around the Milky Way \citep{PolarDarkMatterMilkayway2012}.
{Dark matter can be divided into cold, warm, and hot categories, 
according to the free streaming length (FSL) indicating how far corresponding objects 
moved due to random motions in the early universe. Warm dark matter comprises particles 
with an FSL comparable to the size of a protogalaxy. Predictions based on warm dark 
matter are similar to those for cold dark matter on large scales, but with less small-
scale density perturbations. A conjectured candidate is the sterile neutrino. 
The warm dark matter model is considered to be a better fit to observations. 
However, in both cases, there is currently no experimental detection of dark matter 
particles. Hot dark matter consists of particles, like the neutrino, whose FSL is much 
larger than the size of a protogalaxy. However, current limits on the neutrino masses 
by using cosmic microwave background data induces the result that the neutrinos 
cannot be dark matter.
Here we will follow a different path.}
To explain all these properties, we need a kind of matter which cannot
be compressed, couples to gravity only and is stable. In this paper
we present such a substance: a gravitational soliton 
(in the weak field limit---it will be explained later). 
It is a special very stable structure of the spacetime appearing
in two types, the gravisoliton $\mathcal{S}$ and the anti-gravisoliton
$\mathcal{A}$. Furthermore, there is no coupling to light and therefore
only small acoustic oscillations (caused by possible collisions of
$\mathcal{SS}$ and $\mathcal{AA}$) are allowed. 

{Here we will consider gravisolitons as another component of dark 
matter. Loosely speaking, Gravisolitons are topological defects of the space like
a density wave of space. They behave like matter but with different 
scattering properties.}
At first we will study the properties
of the spacetime which is needed to generate gravitational solitons.
Then we will connect the appearance of gravisolitons with a special
codimension-1 foliation (also known as helical wobble, see \citep{Thu:72}).
In section \ref{sec:How-to-generate} we will discuss how a gravisoliton
can be generated by matter. At the same time, this mechanism uncovers
the topological reason for the appearance of gravisolitons: the underlying
spacetime must admit an exotic smoothness structure. Dark matter was
found by the gravitational lensing effect. Therefore, we will study
the deflection of light rays by the gravisoliton 
in section \ref{sec:Gravitational-lensing-for}.
Interestingly, we will find no difference to ordinary (baryonic) matter. Then we
will discuss a coherent model of spacetime with all necessary properties.
Here, we will see that all relevant spacetime with exotic smoothness
structure are fulfilling these properties. 
{Finally, in section \ref{sec:Measurable-consequences}
we will also discuss the state equation for gravisolitons to be $w=0$, i.e.
no difference to usual matter. The discussion of strong field limit is reserved to
a separate paper.}

\section{Spacetime Properties for the Generation of Gravitational Solitons}

{Dark matter has many properties known for matter itself, 
but it has also some other specific properties, like the 
very weak interaction to usual matter,
which leads to the idea that dark matter consists only 
partly of (particle-like) matter (WIMPS etc.).
Here we will follow this idea, i.e., we will consider dark matter as a
special property of the space. It is just a trial which, nevertheless, 
is aiming toward deciding important question namely whether and how the 
idea can be realized by a spacetime itself. What properties of spacetime are required 
to this end? In the following sections we will describe these properties.
First let us consider the following ansatz for the metric:}
\begin{equation}
ds^{2}=f(t,z)\left(dz^{2}-dt^{2}\right)+g_{ab}(t,z)dx^{a}dx^{b}\label{eq:gravisoliton-metric}
\end{equation}
for $a,b=1,2$ and with the non-diagonal 2-metric $g_{ab}$. The non-diagonality
of $g_{ab}$ is very important, it is responsible for the difference
between non-linear solutions (for non-diagonal $g_{ab}$) and linear
solutions (for diagonal $g_{ab}$). The linear case was first considered
by Einstein and Rosen in 1937 (gravitational cylindrical waves). It
is convenient to introduce the coordinates
\[
\theta=\frac{1}{2}(z-t)\quad\zeta=\frac{1}{2}(z+t)
\]
transforming $dz^{2}-dt^{2}$ into $d\theta\,d\zeta$ and $\alpha=\sqrt{\det(g_{ab})}$
for the determinant. Then the Einstein equation $R_{ab}=0$ is represented
by the matrix equation
\begin{equation}
\frac{\partial}{\partial\theta}\left(\alpha\frac{\partial g}{\partial\zeta}\,g^{-1}\right)+\frac{\partial}{\partial\zeta}\left(\alpha\frac{\partial g}{\partial\theta}\,g^{-1}\right)=0\label{eq:Einstein-matrix-equation}
\end{equation}
and we obtained the additional conditions from the equations $R_{00}=0$,
$R_{33}=0$ and $R_{03}=0$. For the following analysis, we will refer
to the first two chapters of the book \citep{Gravisoliton} where
the solution is presented using the Inverse Scattering Method (ISM).
{The method itself is complicated and cannot be presented here
completely. We will try to give some background here to make the paper as 
self-contained as possible. At first we must explain the wording \emph{weak field 
limit}. A solution of the non-linear equation (\ref{eq:Einstein-matrix-equation})
can be obtained by the ISM as perturbation with respect to a background metric. 
Therefore, this solution is only valid for weakly gravitational fields.
Then the solution is a small perturbation, i.e., we will only obtain a gravisoliton
for weak gravitational fields. Or we will say that the gravisoliton is obtained 
in the weak field limit. The strong field limit will be considered
in a separate paper.}

{Main idea relies on a proper introduction of a background metric. 
Let a particular solution $g_{0}$ serves as a
background metric so that metric $g^{(1)}$ derived from the ISM method is a soliton
solution with respect to $g_{0}$. 
$g_0$ can be chosen as a diagonal metric (for a reason discussed in section \ref{sec:How-to-generate})}
\[
g_{0}=diag\left(\alpha e^{u_{0}},\alpha e^{-u_{0}}\right)
\]
where $\alpha=\sqrt{\det g_{0}}$ and
\[
\frac{\partial^{2}\alpha}{\partial\theta\partial\zeta}=0\qquad\frac{\partial}{\partial\theta}\left(\alpha\frac{\partial u_{0}}{\partial\zeta}\right)+\frac{\partial}{\partial\zeta}\left(\alpha\frac{\partial u_{0}}{\partial\theta}\right)=0
\]
follows from (\ref{eq:Einstein-matrix-equation}). Now we will get
a metric $g$ from the background metric $g_{0}$ by using ISM. 
{Main idea of the ISM is the introduction of two differential operators,
the L-A or Lax pair, so that the eigenfunctions of these operators determine the
new solution out from the old solution. In this case, we started with the constant
diagonal (background) metric. Then the 1-soliton solution is given by
\[
g^{(1)}=\frac{1}{\mu_1\cdot cosh(\rho_1)}\left(
\begin{array}{cc}
(\mu_1^2 e^{\rho_1}+\alpha^2 e^{-\rho_1})e^{u_0}& \alpha^{2}-\mu_{1}^{2} \\
\alpha^{2}-\mu_{1}^{2} & (\mu_1^2 e^{\rho_1}+\alpha^2 e^{-\rho_1})e^{-u_0}
\end{array}
\right)
\]
where $\mu_1,\rho_1$ are integration constants. It is interesting to note that the
off-diagonal elements are the topological charge of the soliton.
Unfortunately, from the physics point of view this solution is difficult to interpret.
Therefore, we will study invariant reparameterizations to study 
the invariant properties of the solution.}
With the notation 
\begin{eqnarray}
\frac{\partial}{\partial\zeta}\left(\ln\left(g_{11}\alpha^{-1}\right)\right)=R_{1}\cos\left(\frac{\gamma+\omega}{2}\right) & \quad & g_{11}\alpha^{-1}\frac{\partial}{\partial\zeta}\left(g_{12}g_{11}^{-1}\right)=R_{1}\sin\left(\frac{\gamma+\omega}{2}\right)\label{eq:equation-system}\\
\frac{\partial}{\partial\theta}\left(\ln\left(g_{11}\alpha^{-1}\right)\right)=R_{2}\cos\left(\frac{\gamma-\omega}{2}\right) & \quad & g_{11}\alpha^{-1}\frac{\partial}{\partial\theta}\left(g_{12}g_{11}^{-1}\right)=R_{2}\sin\left(\frac{\gamma-\omega}{2}\right)\nonumber 
\end{eqnarray}
we have a new parameterization so that $R_{1},R_{2}$ and $\omega$
are an invariant parameterization (not depending on arbitrary linear
transformations of the coordinates $x^{1},x^{2}$) for the metric
$g$ (there is also a system of two equations to determine $\gamma$
which depends only on $R_{1},R_{2}$ and $\omega$). Now we will make
our main assumption:

\[
\alpha=const.\longrightarrow R_{1}=R_{2}=const.
\]
which will be motivated later. Then we obtain the following equation
for the variable $\omega$

\[
\frac{\partial^{2}\omega}{\partial\zeta\partial\theta}=R_{1}R_{2}\sin\omega
\]
which is the famous sine-Gordon equation. This equation admits soliton
solutions of the form (now in coordinates $(z,t)$)

\begin{equation}
\omega(z,t)=4\cdot\arctan\left(e^{mC(z-vt+z_{o})}\right)\label{eq:soliton-solution}
\end{equation}
with velocity $v=-2\alpha\mu_{1}(\alpha^{2}+\mu_{1}^{2})^{-1}$, $C=(1-v^{2})^{-1}$
and mass $m=R_{1}=R_{2}$ which are stable over the time. The constant
$\mu_{1}$ is an integration constant from the ISM. The solution depends
on the sign of $\alpha^{2}-\mu_{1}^{2}$ which is the topological
charge of the soliton. 
{The topological charge is defined as the difference 
$\omega(+\infty,t)-\omega(-\infty,t)$ in units of $2\pi$. Usually, the 
solution (\ref{eq:soliton-solution}) of the sine-Gordon equation describes
a soliton where the sign of $\alpha^{2}-\mu_{1}^{2}$ (topological charge)
determines whether it is a gravisoliton (positive sign) or anti-gravisoliton
(negative sign). Or, the gravisoliton $\mathcal{S}$ has topological
charge $+1$ whereas the anti-gravisoliton $\mathcal{A}$ has $-1$.
The sine-Gordon soliton is a non-linear wave which can be
described by a rotating vector. Then the sign of the topological charge gives the 
direction of the rotation (with respect to the propagation direction).
The difference between $\mathcal{S}$ and $\mathcal{A}$ defines the behavior
with respect to the scattering.
The scattering of solitons like $\mathcal{SS}$ or $\mathcal{AA}$
shows an amazing behavior: there is a repulsive force so that these
solitons bounce back by a collision. In contrast the soliton $\mathcal{SA}$
can be represented by a bound state (a so-called breather solution),
so there is an attractive force in this system. By considering bound
states of solitons (like $\mathcal{SA}$ but technically generated
by a B{\"a}cklund transformation) one obtains 2-gravisolitons which
form only stable stationary solution (including an axisymmetry). 
These 2-solitons in a Minkowski background are equivalent to a
Kerr-NUT or Schwarzschild solution \citep{Gravisoliton}. Then
one obtains an attraction
of a test particle by a 2-gravisoliton (or 2-anti-gravisoliton).} 
Summarizing this approach, we need to construct a spacetime 
\begin{enumerate}
\item with metric (\ref{eq:gravisoliton-metric}): $ds^{2}=f(t,z)\left(dz^{2}-dt^{2}\right)+g_{ab}(t,z)dx^{a}dx^{b}$
where the 2-metric $g_{ab}$ is parameterized by the $t,z-$coordinates only, 
\item the 2-metric $g_{ab}$ is non-diagonal, i.e., $g_{ab}\not=0$ for $a\not=b$ and 
\item $\det(g_{ab})=const.$, i.e., the area of the surface defined
by $g_{ab}$ is fixed. 
\end{enumerate}
How to fulfill these conditions? At first one needs a spacetime which
splits into a (non-trivial, curved) surface and a (conformally flat)
$(1+1)-$dimensional subspace (first condition). The surface is locally
generated by two independent vector fields (the dual to $dx^{1},dx^{2}$)
which are never orthogonal to each other (second condition). Furthermore
this surface has a finite volume (third condition). The first and
third conditions can be simply realized for instance by spaces like
$\mathbb{R}^{1,1}\times S$ for every compact closed surface $S$.
The problem is only with the second condition. In principle, we need a complicated
codimension-1 foliation of a 3-dimensional manifold (formed by $x^{1},x^{2},z$)
so that the three generating vector fields are never orthogonal to
each other. In the next section we will construct this foliation for
the 3-sphere but also for all other compact 3-manifolds.

\section{A codimension-1 Foliation Realizing the Conditions for Gravitational
Solitons}\label{topo-prelimenaries}
{In the previous section we discussed gravitational solitons, called 
gravisolitons, which are connected with the metric (\ref{eq:gravisoliton-metric}). Main 
characteristics of this metric is the appearance of a non-diagonal 2-metric $g_{ab}$. 
Then the non-diagonal term of the 2-metric can be interpreted as non-orthogonal vector 
fields in this space. With other words, we must construct a space or spatial 
subspace of the spacetime which is generated by non-orthogonal vector fields. A direct 
way would be the foliation of the space by surfaces described by the 2-metric $g_{ab}$. 
Does such foliation exist? And if it exists, what are conditions guaranteeing this 
which must be implied on the foliation and the spacetime? \\
First, we will give a flavor of the main idea without the technical
details. Our task is now to construct a foliation of a 3-manifold into surfaces so that
the three generating vector fields are never orthogonal to each other. This problem was 
solved by Thurston \cite{Thu:72} who constructed a codimension-1 foliation of the 3-
sphere which can be easily extended to all 3-manifolds. The details of the complicated 
construction are not important for the following discussion and can be found in 
\cite{AsselmeyerMaluga2016}. Main idea is the usage of the group 
$PSL(2,\mathbb{R})$. Then the vector fields of the foliation are the left-invariant 
vector fields of the Lie algebra of this group. Furthermore, we will construct an 
invariant of the foliation (Godbillon-Vey invariant) which is directly connected to the 
non-orthogonal vector fields. Differently expressed: the vector fields are non-
orthogonal if this invariant is non-zero. The reader not interested in the technical 
details of this construction can switch to the next section keeping in mind that there 
is such a foliation with the suitable properties.}

A foliation $(M,F)$ of a manifold $M$ is an integrable
Sub-bundle $F\subset TM$ of the tangent bundle $TM$. The leaves $L$
of the foliation $(M,F)$ are the maximal connected submanifolds $L\subset M$
with $T_{x}L=F_{x}$ for all $x\in L$. The number $\dim M-\dim L$
is the codimension of the foliation. A codimension-1 foliation on
a 3-manifold $M$ can be constructed by a smooth 1-form $\omega$
fulfilling the integrability condition $d\omega\wedge\omega=0$.  This construction
can be found in \cite{AsselmeyerMaluga2016}. In \citep{ReinhartWood1973}
the authors analyzed the local structure of the foliation: the three
vector fields (forming the frame of the foliation) are not orthogonal
to each other. This foliation is constructed by using the group
$PSL(2,\mathbb{R})$, the isometry group of the hyperbolic plane $\mathbb{H}^{2}$.
Let $T,N,Z$ be the frame formed by three vector field dual to the
one-forms $\omega,\eta,\xi$ as a basis for the tangent bundle $TM$
(remember that the tangent bundle of every 3-manifold is trivial,
see \citep{MSt:74}). Now one considers the vector fields as left-invariant
vector fields of the group $PSL(2,\mathbb{R})$ (forming the basis
of the Lie algebra) fulfilling the commutator relations $[Z,N]=Z$,
$[N,T]=T$ and $[Z,T]=N$. As shown in \citep{ReinhartWood1973},
the vector fields are not orthogonal to each other leading to a (local)
non-zero torsion $\tau\not=0$. For codimension-1 foliations of 3-manifolds,
there exists the Godbillon-Vey invariant which we are going to
describe now. Recall that codimension-1 foliation on $M$ is defined by a
$PSL(2,\mathbb{R})-$invariant integrable one-form $\omega$ ($d\omega\wedge\omega=0$)
and we define another one-form $\eta$ by $d\omega=-\eta\wedge\omega$.
Then the Godbillon-Vey class $gv=\eta\wedge d\eta$ is a closed form
(by using the integrability) which is not exact. Furthermore, the integral

\begin{equation}
GV(M)=\intop_{M}\eta\wedge d\eta\label{eq:godbillon-vey-inv}
\end{equation}
is a topological invariant of the foliation, known as Godbillon-Vey
number $GV(M)$. From the physics point of view, it is the abelian
Chern-Simons functional. The mathematical construction of the foliation
can be found in \cite{Thu:72,Tamura1992} including a global calculation
of the Godbillon-Vey invariant. However, here we will concentrate on the
local expression for the Godbillon-Vey invariant. Let $\kappa,\tau$
be the curvature and torsion of a normal curve to the foliation, respectively.
Furthermore, let $T,N,Z$ be the frame formed by three vector field
dual to the one-forms $\omega,\eta,\xi$ and let $l_{T}$ be the second
fundamental form of the leaf. Then the Godbillon-Vey class is locally
given by

\begin{equation}
\eta\wedge d\eta=\kappa^{2}\left(\tau+l_{T}(N,Z)\right)\omega\wedge\eta\wedge\xi\label{eq:GV-locally}
\end{equation}
where $\tau\not=0$ for $PSL(2,\mathbb{R})$ invariant foliations
i.e., $[Z,N]=Z$, $[N,T]=T$ and $[Z,T]=N$. Therefore at least two
vector fields are not orthogonal to each other and the corresponding
metric $g_{ab}$ is not diagonal. Finally, let $M$ be a compact 3-manifold with codimension-1 foliation
of non-zero Godbillon-Vey number $GV(M)\not=0$. Then the spacetime
$M\times\mathbb{R}$ (or $M\times[0,1]$ for finite time) has all
properties to admit gravitational solitons (in the weak field limit).

\section{ How to Generate Gravitational Solitons\label{sec:How-to-generate}}

{In the previous section we connected the gravitational soliton with 
a foliation of the 3-space. The result can be expressed by the simple statement: if the 
space admits a codimension-1 foliation with non-zero Godbillon-Vey invariant 
(\ref{eq:godbillon-vey-inv}), locally given by (\ref{eq:GV-locally}), then, there is a 
gravitational soliton (gravisoliton) in space described by the metric 
(\ref{eq:gravisoliton-metric}). 
Therefore, if we can get a non-zero Godbillon invariant then a gravisoliton is 
generated. Starting point is a general but arbitrary foliation of the space into 
surfaces. In \citep{AsselmeyerBrans2015}, we described a strong relation (using
the work of \citep{Friedrich1998}) between a surface and a spinor $\psi$. We will give 
a different meaning to the Godbillon-Vey invariant. In principle, this invariant is 
generated by a 3-form $\eta\wedge d\eta$ with the real 1-form $\eta$.  
Now we interpret $\eta$ as a $U(1)$ gauge field $A$ which is an imaginary valued 1-form 
i.e., $A=i\eta$. Then we couple this gauge field $\eta$ to the spinor $\psi$. The 
Godbillon-Vey invariant is also known as abelian Chern-Simons form. 
Then the action of a spinor coupled to the gauge field given by the Chern-Simons form 
is a dynamical theory to generate a foliation of the space (see the Dirac-Chern-Simons 
functional below). The foliation is defined by the gradient flow.
However, the gradient flow is equivalent to the Seiberg-Witten equations, a theory to detect 
different smoothness structures on 4-manifolds.
A non-trivial solution of this gradient system is given by $\psi\not=0,\eta\not=0$, 
which is equivalent to the existence of a non-trivial Godbillon-Vey invariant which 
means that a gravisoliton is being generated. 
However, this solution gives also a non-trivial solution of the Seiberg-Witten equation for 
the 4-dimensional spacetime. This solution is a necessary condition for an exotic 
smoothness structure of the spacetime. A smoothness structure is the maximal smooth 
atlas of a manifold, i.e., the collection of smooth charts covering the manifold. In 
principle, the smoothness structure is the main topological information about the 
smooth manifold. Loosely speaking, all properties of differentiable quantities are 
determined by the smoothness structure, even one cannot define a differential equation 
on a manifold without fixing the smoothness structure. In dimension 4 there are 
manifolds which are homeomorphic to each other but not diffeomorphic, i.e., these 
manifolds differ by the smoothness structure. Main results of this section are the 
connection between the existence of gravisolitons in space and the smoothness structure 
of the spacetime. }

{Now we go into the details of this construction. In 
\citep{AsselmeyerBrans2015} we presented a geometrical/topological model for matter. In 
general, for a manifold $M$ with boundary $\partial M=\Sigma$
one has the action (see \cite{GibHaw1977}) 
\[
S_{EH}(M)=\intop_{M}R\sqrt{g}\:d^{4}x+\intop_{\Sigma}H\,\sqrt{h}\,d^{3}x
\]
where $H$ is the mean curvature of the boundary with metric $h$.
For the following discussion, we consider the boundary term
\begin{equation}
S_{EH}(\Sigma)=\intop_{\Sigma}H\,\sqrt{h}\,d^{3}x\label{eq:action fermi}
\end{equation}
along the boundary $\Sigma$ (a 3-manifold). Following \cite{AsselmeyerBrans2015},
the action (\ref{eq:action fermi}) over a 3-manifold $\Sigma$ is
equivalent to the Dirac action of a spinor over $\Sigma$. Interestingly, this relation 
is induced by a stronger relation between surfaces and spinors (see 
\citep{AsselmeyerRose2012} for a complete discussion). }
Main result of \citep{AsselmeyerBrans2015} is the following relation
between the corresponding Dirac operators

\begin{equation}
D^{M}\Phi=D^{\Sigma}\psi-H\psi\label{eq:relation-Dirac-3D-4D}
\end{equation}
where $D^{\Sigma}$ or $D^{M}$ denote the Dirac operators on the
3-manifold $\Sigma$ or 4-manifold $M$, respectively. 
{Using the work of \cite{Friedrich1998},
the spinor $\phi$ directly defines the embedding (via an integral
representation) of the 3-manifold. Then the restricted spinor $\Phi|_{\Sigma}=\phi$
is parallel transported along the normal vector and $\Phi$ is constant
along the normal direction. However, then the spinor $\Phi$ must fulfill
\begin{equation}
D^{M}\Phi=0\label{eq:Dirac-equation-4D}
\end{equation}
i.e., $\Phi$ is a parallel spinor. Finally, we get 
\begin{equation}
D^{\Sigma}\psi=H\psi\label{eq:Dirac3D-mean-curvature}
\end{equation}
and 
\begin{equation}
\intop_{\Sigma}H\,\sqrt{h}\,d^{3}x=\intop_{\Sigma}\bar{\psi}\,D^{\Sigma}\psi\,\sqrt{h}
d^{3}x\label{eq:relation-mean-curvature-action-to-dirac-action}
\end{equation}
Therefore, we get a direct relation between spinors and geometry. Now we will use this 
relation to get a connection to the foliation.}

For every codimension-1 foliation, it is known: If the 1-form $\omega$ 
defines the leaf via the equation $\omega=const.$ then
the dual of the 1-form $\eta$ defines a vector normal to the leaf.
However, then the spinor $\Phi$ (representing the leaf) is constant along
the normal direction fulfilling the relation (\ref{eq:Dirac-equation-4D})
i.e., $\Phi$ is a parallel spinor. Furthermore, in \citep{AsselmeyerBrans2015}
we showed that a fermion is given by a knot complement admitting a
hyperbolic structure. Then calculations in \citep{Asselmeyermaluga2019}
implied that the particular knot is only important for the dynamical
state (like the energy or momentum) but not for charges, flavors etc. 

Now we will use this relation between spinors, surfaces and the foliation
to reinterpret the Dirac equation $D^{M}\Phi=0$ along the normal
direction as covariant constant spinor
\[
D_{\eta}^{\Sigma}\psi=0
\]
where the one-form $\eta$ is interpreted as abelian gauge field (with structure group 
$U(1)$.
{By definition, the covariant constant 1-form $\omega$
\[
D_{\eta}\omega=d\omega+\eta\wedge\omega=0
\]
defines the foliation}, because the integrability condition $\omega\wedge d\omega=0$
is automatically fulfilled. Then we reinterpret the Godbillon-Vey
invariant $gv=\eta\wedge d\eta$ of the foliation as abelian Chern-Simons
form for the abelian gauge field $\eta$. 
{Now we use the relation between the spinor and the surface (as given by 
$\omega=const.$) to define a similar relation as $D_{\eta}\omega=0$: the spinor $\psi$ 
is covariant constant with respect to $\eta$ or
\[ D_{\eta}^{\Sigma}\psi=0 \,.\] 
To take the special properties of the foliation into account, we must consider 
a non-zero Godbillon-Vey invariant at the same time.}
That means that for a fixed
foliation, the coupling between the abelian gauge field $\eta$ and
the spinor $\psi$ to the Dirac-Chern-Simons action functional 
\[
S_{DCS}=\intop_{\Sigma}\left(\bar{\psi}\,D_{\eta}^{\Sigma}\psi\,\sqrt{h}d^{3}x+\eta
\wedge d\eta\right)
\]
on the 3-manifold is constant with the critical points at the solution
\[
D_{\eta}^{\Sigma}\psi=0\quad d\eta=\tau(\psi,\psi)
\]
where $\tau(\psi,\psi)$ is the unique quadratic form for the spinors
locally given by $\bar{\psi}\gamma^{\mu}\psi$. If one has a spacetime
$\Sigma\times I$, then we consider the solution which is translationally
invariant. Alternatively, it is a spacetime with foliation induced
by the foliation of $\Sigma$ extended by a translation. Mathematically
we must consider the gradient flow
\begin{eqnarray*}
\frac{d}{dt}\eta & = & d\eta-\tau(\psi,\psi)\\
\frac{d}{dt}\psi & = & D_{\eta}^{\Sigma}\psi
\end{eqnarray*}
{From every solution of this gradient system, one can construct the
corresponding foliation with Godbillon-Vey invariant. It is interesting to note
that a solution with vanishing Godbillon-Vey invariant leads also to vanishing spinor.}
In \citep{MorSzaTau:96,MorSzaTau:97} this system was shown to be
equivalent to the Seiberg-Witten equation for $\Sigma\times I$ by
using an appropriate choice of the $Spin_{C}$ structure . Then we
have the result that a non-trivial foliation together with the existence
of fermions induce a non-trivial solution of the gradient system which
results in a non-trivial solution of the Seiberg-Witten equations.
However, this non-trivial solution (i.e., $\psi\not=0,\eta\not=0$) is a
necessary condition for the existence of an exotic smoothness structure
(but not sufficient). 

{Here we get an unexpected relation between the existence of 
gravisolitons and topological (or better differential-topological) properties of the 
spacetime. 
This property of the spacetime is called exotic smoothness in mathematics (see the book 
\citep{Asselmeyer2007} for an overview).

Again, exotic smoothness denotes a different (non-diffeomorphic)
smooth atlas of a 4-manifold. Two 4-manifolds with different smoothness
structures are topologically equivalent (homeomorphic) but smoothly
different. From the physics point of view, two non-diffeomorphic spacetimes
are different if they represent (physically) different systems.

Finally, we are concluding that the foliation above is
connected with the smoothness properties of the spacetime. Furthermore, if one starts 
choosing an exotic smoothness structure instead of the standard one for the spacetime 
then, there gravisolitons must exist automatically. By using the relation between 
spinors and surfaces, there is a natural coupling between a spinor and the 
gravisoliton, or the gravisoliton is generated by a spinor (electrically charged). The 
description here is only the first sign of how the appearance of gravisolitons could be 
understood in terms of physical spinor fields. More work is needed to understand it 
completely.}

\section{Gravitational Lensing for Gravitational Solitons}
\label{sec:Gravitational-lensing-for}

{In this section, we will investigate the gravitational lensing property 
of the gravitational solitons. To study the effect of these solitons on the propagation 
of light, we will introduce the optical metric and compare it with the metric of the 
gravitational soliton. Unexpectedly, we will obtain the same effect as for usual 
matter. Therefore, one cannot see a difference between the gravitational soliton and 
matter by using gravitational lensing. This section is important to understand that 
gravitational solitons cannot be distinguished from matter by gravitational lensing.}

Dark matter was indirectly recognized by gravitational lensing. Therefore
we are going to study the deflection of light induced by gravitational
solitons. The idea is to introduce a metric, the optical metric, which describes this 
deflection of light.

{The optical metric with the line element $ds_{opt}^2$
in Schwarzschild coordinates $(t,r,\vartheta,\phi)$ is usually given by
\[
ds_{opt}^{2}=-exp\left(2A(r)\right)dt^{2}+exp\left(2B(r)\right)dr^{2}+r^{2}\left(d
\vartheta^{2}+sin^{2}\vartheta\,d\phi^{2}\right)
\]
with some functions $A(r),B(r)$. To determine these two functions, one must put it in 
Einstein's field equations. Here one uses an energy-momentum tensor $T_{\mu\nu}
=diag(\rho,p,p,p)$ in the local flat metric, where $\rho = \rho(r)$ denotes the density 
and $p = p(r)$ the pressure of the lens model. 
Then one obtains
\[
exp\left(-2B(r)\right)=1-\frac{2\mu(r)}{r}
\]
with
\[
\mu(r)=4\pi G\intop_{0}^{r}\rho(r')r'^{2}dr'
\]
where $\rho(r)$ denotes the energy density.} 
However, the solution (\ref{eq:soliton-solution})
implied that the $dr^{2}$ part of the corresponding gravitational
soliton solution is in a good approximation constant, so that we get
\[
exp\left(-2B(r)\right)=1-\frac{2\mu(r)}{r}=const.
\]
However, then we will obtain
\[
\mu(r)\sim r
\]
and the energy density of a gravitational soliton is given by
\[
\rho(r)\sim\frac{1}{r^{2}}
\]
i.e., we obtain the same gravitational lensing effect as the singular
isothermal sphere serving as a lensing model for galaxies. In 
\citep{GibbonsWerner2008},
the deflection angle was determined from the above relation.
{Let us explain briefly how the deflection angle in the solitonic model 
is obtained.
Let $\sigma^{2}$ be the amplitude of the soliton, then we set
\[
\rho(r)=\frac{\sigma^{2}}{2\pi Gr^{2}}
\]
implying $\mu(r)=2\sigma^{2} r$. This is known as the solution of the Tolman-
Oppenheimer-Volkoff equation in the nonrelativistic limit (i.e., first order in $
\sigma^2$). 
The Gaussian curvature vanishes for $r > 0$ in this limit, so that the equatorial plane 
in the optical metric is a cone with a singular vertex at $r = 0$. 
The deficit angle of the conical optical metric is the deflection angle $\delta$.
Thus, for our model the deflection angle is given by
\[
\delta=4\pi\sigma^{2}
\]
in agreement with the gravitational lensing of galaxies. 
It is interesting to note that this lensing is similar to the lensing of a cosmic 
string.}
Finally, gravitational solitons have the same effect as usual matter and gravitational 
lensing cannot distinguish between the gravitational soliton and matter.

\section{A Coherent Model for a Spacetime with These Properties}
\label{sec:A-coherent-model}

{In the section \ref{sec:How-to-generate}, we found an unexpected 
relation between gravisolitons and (differential-)topological properties of the 
spacetime. In particular, the existence of gravisolitons is connected with a class of 
complicated foliations (with non-trivial Godbillon-Vey invariant) leading to a
coupling of the gravisoliton to a spinor. Then the last relation (spinor-gravisoliton 
coupling) can be fulfilled for a class of spacetimes, known as spacetime with exotic 
smoothness structure. In this section, we will construct a suitable spacetime and study 
its properties. Finally, we will see that for this class of spacetimes we obtain the
remaining conditions which are needed to construct the gravisoliton (i.e., $\alpha=
\det(g)=const.$ and the non-diagonality of the 2-metric).}

{In a series of papers we have described the class of spacetimes with 
exotic smoothness structure. The influence of the smoothness structure on the
topological properties of the spacetime is tremendous. Let us consider the spacetime
given by the smooth manifold $S^{3}\times\mathbb{R}$. In the standard smoothness 
structure on $S^{3}\times\mathbb{R}$ one has a global foliation of the spacetime into 
slides $S^{3}\times\{t\}$ for every 
$t\in\mathbb{R}$ (or by copies of $S^{3}$). In contrast to that, the 4-spacetime $S^{3}
\times\mathbb{R}$ with exotic smoothness structure is given by a sequence $\ldots\to
\Sigma_{1}\to\Sigma_{2}\to\ldots$
of topological transitions where all $\Sigma_{i}$ have the same homology
like $S^{2}\times S^1$. In the limit of infinite transitions, one gets a manifold which 
is topologically equivalent to the 3-sphere. We remark that this is a general
rule for all exotic versions of spacetimes $N\times\mathbb{R}$ (with
$N$ a compact 3-manifold). To understand the general structure of the $\Sigma_n$, we 
must understand the general decomposition of 3-manifolds into pieces.}

{Every 3-manifold can be decomposed into prime
3-manifolds according to 
\[
K_{1}\#\cdots\#K_{n_{1}}\#_{n_{2}}S^{1}\times S^{2}\#_{n_{3}}S^{3}/\Gamma
\]
where $\Gamma$ is a finite subgroup of $SO(4)$. Every (irreducible)
3-manifold $K_{i}$ can be also split into a hyperbolic and a graph
manifold.  As shown in \citep{AsselmeyerRose2012}
and further extended in \citep{AsselmeyerBrans2015,Asselmeyermaluga2019},
the 3-manifolds $K_{i}$ are the matter part. It consists of
hyperbolic 3-manifolds as fermions and graph manifolds (torus bundles)
for the interaction (gauge bosons). The further details of this model
are not important here, but we want to call the reader's attention
to the sum operation $\#$. By definition we have for the connecting
sum $K_1\#K_{2}$ the expression
\[
K_1\#K_{2}=\left(K_1\setminus D^{3}\right)\cup_{S^{2}}\left(S^{2}\times[0,1]\right)
\cup_{S^{2}}\left(K_{2}\setminus D^{3}\right)
\]
with the tube $S^{2}\times[0,1]$ to identify $K_1\setminus D^{3}$
and $K_{2}\setminus D^{3}$ along its common boundary $S^{2}$. Therefore
we obtained an additional manifold $S^{2}\times[0,1]$, called sphere bundle, 
between the matter $K_{i}$. The corresponding
spacetime version of these sphere bundles admit the topology $S^{2}\times[0,1]^{2}$.
Obviously, we have the following properties: 
\begin{enumerate}
\item $S^{2}\times[0,1]^{2}$ admits a Lorentz metric with coordinates $(x^{1},x^{2})$
for the sphere $S^{2}$ and $(z,t)$ for $[0,1]^{2}$ and 
\item the line element is given by
\[
ds^{2}=f(t,z)\left(dz^{2}-dt^{2}\right)+g_{ab}(t,z)dx^{a}dx^{b}
\]
in agreement with (\ref{eq:gravisoliton-metric}). 
\end{enumerate}
The next two properties of the spacetime
are directly related to the exotic smoothness structure on it. 
However, from the point of view of geometry, exotic smoothness is connected with
hyperbolic geometry (see for instance \citep{Lebrun96,Lebrun98} and the result stating 
that 4-manifolds with positive scalar curvature in the standard smoothness structure
fail to be paired with hyperbolic geometry). In
\citep{AsselmeyerKrol2014} we analyzed this situation and explained the reason:
exotic smoothness structures enforce the appearance of saddle points
(in the Morse-theoretic sense, i.e., non-canceling 2-/3-handle pairs)
carrying a hyperbolic geometry. The corresponding analysis was extended
in \citep{AsselmeyerKrol2018c,AsselmeyerMaluga2018d}. This fact has
two important consequences: hyperbolic 3- and 4-manifolds (with finite
volume, i.e., the neighborhood of the saddle points) have the property
called {\em Mostow rigidity}. As shown by Mostow \citep{Mos:68}
and extended by Prasad \citep{Prasad:1973}, every hyperbolic $n-$manifold
$n>2$ (with finite volume) has the following property: \emph{Every diffeomorphism
(especially every conformal transformation) of a hyperbolic $n-$manifold
(with finite volume) for $n>2$ is induced by an isometry.} Therefore, one cannot
scale a hyperbolic 3- and 4-manifold. Then the volume $vol(\:)$ and
the curvature (both combined into the Chern-Simons invariant) are
topological invariants. Secondly, the foliation of this hyperbolic
4-manifold part admits a non-trivial Chern-Simons invariant and 
also a non-trivial  Godbillon-Vey invariant (foliations of type III) 
\citep{AsselmeyerKrol2009}. Both consequences imply
the following properties:}
\addtocounter{enumi}{2} 
\begin{enumerate}
\item The background for the sphere bundle has a constant scale-invariant
volume (by Mostow rigidity) from the matter, i.e., $\det(g)=const.$ 
\item The metric $g$ is non-diagonal (otherwise the Godbillon-Vey invariant
of the foliation vanishes, see \citep{HurKat:84} and sections \ref{topo-prelimenaries}, 
\ref{sec:How-to-generate}). 
\end{enumerate}
All four properties together are the prerequisites to obtain the model
of the gravisoliton as dark matter.

\section{The Equation of State for Gravisolitons}\label{sec:Measurable-consequences}

{In the astrophysical context, the scaling behavior of 
the energy density $\rho(a)$ depends on the scale factor $a(t)$. Now we
consider the fluid equation
\[
\dot{\rho}+\frac{3\dot{a}}{a}(\rho +p)=0
\] 
and the equation of state $p=w\cdot\rho$. Usually, one assumes a constant 
$w$ so that one gets
\[ \rho\sim a^{-3(1+w)} \]
For gravisolitons, the energy $E$ is constant.
In the previous section we discussed that the volume of the sphere $S^2$ in the
sphere bundle $S^2\times [0,1]$ (representing the gravisoliton) is fixed. 
Furthermore, we know from exotic smoothness structure that this sphere bundle
is the equator region of a 3-sphere which carries a hyperbolic structure.
By Mostow rigidity (see section \ref{sec:A-coherent-model}), the volume of 
$S^2\times [0,1]$ is also fixed, i.e., gravisolitons do not scale with the 
expansion of the universe. Then we obtain 
\[
\rho_{soliton}=\sum_i^N \frac{\lambda E_i}{V}
\]   
where $N$ is the number of gravisolitons in our cosmic horizon, 
$\lambda$ is the linear density of gravisolitons, and $E_i$ is the energy 
of each gravisoliton. Then we can get the dependence of $\rho$ on the scale $a$ 
\[
\rho_{soliton}(a)\sim \frac{1}{a^3}=a^{-3}
\]
and then
\[
3=3(1+w_{soliton})\Longrightarrow w_{soliton}=0
\]
}
Then we have the following consequences: 
\begin{itemize}
\item (i) The state equation of the gravisoliton (as dark matter) is 
$p=0$ or $w=0$. 
\item (ii) Gravisolitons interact only via gravity with matter by an attractive
force. 
\item (iii) {Gravisolitons have two possible scattering 
behaviors: two gravisolitons scatter so that it looks repulsively. 
In contrast, gravi- and anti-gravisolitons scatter so that it looks attractively. }
\end{itemize}
{In \citep{DarkMatterEquation2011}, the state equation for
dark matter was investigated by using the gravitational lensing. 
For two clusters, the Coma and the CL0024 cluster (see Fig. 2 in 
\citep{DarkMatterEquation2011}), the variation of the equation of state parameter
$w$ over the radius of the cluster was studied. In contrast to the expected value
at $w=0$, the both curves are located at the value $w=-1/3$ but the error bars are 
large enough to get $w=0$ also. Newer measurements \citep{DarkMatterEquation2014}
got the value $w=0$. In \citep{DarkMatterEquation2018} the equation of state for
dark matter was studied. In particular, the inverse cosmic volume law for 
Dark Matter was tested by allowing its equation of state to vary independently 
for different times values in the cosmic history (i.e., eight redshift bins 
in the range $z=10^5$ and $z=0$). Again, the equation of state $w=0$ was also 
confirmed during the cosmic evolution.
According to consequence (i) above, our results agree with
these measurements. Therefore, the detection of gravisolitons is rather complicate.\\
However, what is different for gravisolitons? From the mathematics point of view, the 
corresponding foliation leading to 
gravisolitons (see section \ref{topo-prelimenaries}) admits torsion. Here 
we expect that the polarization properties of the light after the gravitational 
lensing must be different. We conjecture that the light must be circularly polarized.
However, what about the strong field limit? One of the main results is the close 
relation between the smoothness properties and the existence of gravisolitons.
As discussed in our previous work, the space admits a fractal structure.
Gravisolitons will follow this structure and the distribution of dark matter
looks like a fractal which will be shown in a forthcoming paper.}

%
%
%

\acknowledgments{We thank the anonymous referees for helpful remarks. In the course of the revision
we found an error which we correct now. Furthermore, we acknowledge the critical
remarks and questions which increased the readability of this paper.}


%


\begin{thebibliography}{99}
\providecommand{\natexlab}[1]{#1}

\bibitem[Zwicky(1933)]{Zwicky1933}
Zwicky, F.
\newblock Die {R}otverschiebung Von Extragalaktischen {N}ebeln.
\newblock {\em Helvetica Physica Acta} {\bf 1933}, {\em {6}},~110--127.

\def\apj{Astrophys. J.}
\bibitem[Zwicky(1937)]{Zwicky1937}
Zwicky, F.
\newblock On the Masses of Nebulae and of Clusters of Nebulae.
\newblock {\em \apj} {\bf 1937}, {\em { 86}},~217--246.


\bibitem[Rubin \em{et~al.}(1980)Rubin, Thonnard, and Ford]{Rubin1980}
Rubin, V.; Thonnard, W.K.J.; Ford, N.
\newblock Rotational Properties of 21 Sc Galaxies with a Large Range of
  Luminosities and Radii from {NGC} 4605 (R = 4kpc) to {UGC} 2885 (R = 122kpc).
\newblock {\em \apj} {\bf 1980}, {\em 238},~471--487.

\bibitem[Komatsu \em{et~al.}(2011)Komatsu, Smith, Dunkley, Bennett, Gold,
  Hinshaw, Jarosik, Larson, Nolta, Page, Spergel, Halpern, Hill, Kogut, Limon,
  Meyer, Odegard, Tucker, Weiland, Wollack, and Wright]{WMAP-7-years}
Komatsu, E.; Smith, K.M.; Dunkley, J.; Bennett, C.L.; Gold, B.; Hinshaw, G.;
  Jarosik, N.; Larson, D.; Nolta, M.R.; Page, L.; et al.
\newblock Seven-year {Wilkinson Microwave Anisotropy Probe (WMAP)}
  Observations: Cosmological Interpretation.
\newblock {\em Astrophys. J. Suppl.} {\bf 2011}, {\em 192},~18.

\bibitem[{Ade et. al.}(2013)]{PlanckCosmoParameters2013}
{Ade et. al.}, P.
\newblock Planck 2013 Results. {XVI}. Cosmological Parameters.
\newblock arXiv:1303.5076[astro-ph.CO].

\bibitem[Metz \em{et~al.}(2009)Metz, Kroupa, Theis, Hensler, and
  Jerjen]{MilkaywayDwarfGalaxies2009}
Metz, M.; Kroupa, P.; Theis, C.; Hensler, G.; Jerjen, H.
\newblock Did the Milkay Way Dwarf Satellites Enter the Halo as a Group?
\newblock {\em \apj} {\bf 2009}, {\em {
  697}},~269--274.

\bibitem[Moni~Bidin \em{et~al.}(2012)Moni~Bidin, M{\'e}ndez, and
  Smith]{DarkMatterSunsystem2012}
Moni~Bidin, C.~AndCarraro, G.; M{\'e}ndez, R.; Smith, R.
\newblock Kinematical and Chemical Vertical Structure of the Galactic Thick
  Disk {II}. A Lack of Dark Matter in the Solar Neighborhood.
\newblock {\em \apj} {\bf 2012}, {\em { 751}},~30.

\bibitem[Pawlowski \em{et~al.}(2012)Pawlowski, Pflamm-Altenburg, and
  Kroupa]{PolarDarkMatterMilkayway2012}
Pawlowski, M.; Pflamm-Altenburg, J.; Kroupa, P.
\newblock The {VPOS}: A Vast Polar Structure of Satellite Galaxies, Globular
  Clusters and Streams Around the Milky Way.
\newblock {\em Mon. Not. R. Astron. Soc.} {\bf 2012}, {\em {
  423}},~1109--1126.

\bibitem[Thurston(1972)]{Thu:72}
Thurston, W.
\newblock Noncobordant Foliations of {$S^3$}.
\newblock {\em Bull. Am. Math. Soc.} {\bf 1972}, {\em 78},~511--514.

\bibitem[Belinski and Verdaguer(2004)]{Gravisoliton}
Belinski, V.; Verdaguer, E.
\newblock {\em Gravitational Solitons}; Cambridge University Press: Cambridge,
  2004.

\bibitem[Asselmeyer-Maluga(2016)]{AsselmeyerMaluga2016}
Asselmeyer-Maluga, T.
\newblock Smooth Quantum Gravity: Exotic Smoothness and Quantum Gravity. In
  {\em At the Frontiers of Spacetime: Scalar-Tensor Theory, Bell's Inequality,
  Mach's Principle, Exotic Smoothness}; Asselmeyer-Maluga, T., Ed.; Springer:
  Switzerland,  2016.
\newblock in honor of Carl Brans's 80th birthday, arXiv:1601.06436.

\bibitem[Reinhart and Wood(1973)]{ReinhartWood1973}
Reinhart, B.; Wood, J.
\newblock A Metric Formula for the {G}odbillon-{V}ey Invariant for Foliations.
\newblock {\em Proc. AMS} {\bf 1973}, {\em {\bf 38}},~427--430.

\bibitem[Milnor and Stasheff(1974)]{MSt:74}
Milnor, J.; Stasheff, J.
\newblock {\em Characteristic Classes}; Ann. Math. Studies,76, Princeton Univ.
  Press: Princeton, N.J.,  1974.

\bibitem[Tamura(1992)]{Tamura1992}
Tamura, I.
\newblock {\em Topology of Foliations: An Introduction}; Translations of Math.
  Monographs Vol. 97, AMS: Providence,  1992.

\bibitem[Asselmeyer-Maluga and Brans(2015)]{AsselmeyerBrans2015}
Asselmeyer-Maluga, T.; Brans, C.
\newblock How to Include Fermions Into General Relativity by Exotic Smoothness.
\newblock {\em Gen. Relativ. Grav.} {\bf 2015}, {\em {\bf 47}},~30.

\bibitem[Friedrich(1998)]{Friedrich1998}
Friedrich, T.
\newblock On the Spinor Representation of Surfaces in Euclidean 3-Space.
\newblock {\em J. Geom. and Phys.} {\bf 1998}, {\em 28},~143--157.


\bibitem[Gibbons and Hawking(1977)]{GibHaw1977}
Gibbons, G.; Hawking, S.
\newblock Action Integrals and Partition Functions in Quantum Gravity.
\newblock {\em Phys. Rev. D} {\bf 1977}, {\em 15},~2752--2756.

\bibitem[Asselmeyer-Maluga and Ros{\'e}(2012)]{AsselmeyerRose2012}
Asselmeyer-Maluga, T.; Ros{\'e}, H.
\newblock On the Geometrization of Matter by Exotic Smoothness.
\newblock {\em Gen. Rel. Grav.} {\bf 2012}, {\em {\bf 44}},~2825 -- 2856.


\bibitem[Asselmeyer-Maluga(2019)]{Asselmeyermaluga2019}
Asselmeyer-Maluga, T.
\newblock Braids, 3-Manifolds, Elementary Particles: Number Theory and Symmetry
  in Particle Physics.
\newblock {\em Symmetry} {\bf 2019}, {\em 11(10)},~1298.


\bibitem[Morgan \em{et~al.}(1996)Morgan, Szabo, and Taubes]{MorSzaTau:96}
Morgan, J.; Szabo, Z.; Taubes, C.
\newblock A Product formula for the Seiberg-Witten invariants and the
  Generalized Thom Conjecture.
\newblock {\em J. Diff. Geom.} {\bf 1996}, {\em {\bf 44}},~706--788.

\bibitem[Morgan \em{et~al.}(1997)Morgan, Szabo, and Taubes]{MorSzaTau:97}
Morgan, J.; Szabo, Z.; Taubes, C.
\newblock Product formulas along $T^3$ for Seiberg-Witten invariants.
\newblock {\em Mathematical Research Letters} {\bf 1997}, {\em {\bf
  4}},~915--929.

\bibitem[Asselmeyer-Maluga and Brans(2007)]{Asselmeyer2007}
Asselmeyer-Maluga, T.; Brans, C.
\newblock {\em Exotic {S}moothness and {P}hysics}; WorldScientific Publ.:
  Singapore,  2007.

\bibitem[Gibbons and Werner(2008)]{GibbonsWerner2008}
Gibbons, G.; Werner, M.
\newblock Applications of the Gauss-Bonnet theorem to gravitational lensing.
\newblock {\em Class. Quant. Grav.} {\bf 2008}, {\em {\bf 25}},~235009.


\bibitem[LeBrun(1996)]{Lebrun96}
LeBrun, C.
\newblock Four-Manifolds Without Einstein Metrics.
\newblock {\em Math. Res. Lett.} {\bf 1996}, {\em 3},~133--147.

\bibitem[LeBrun(1998)]{Lebrun98}
LeBrun, C.
\newblock Weyl Curvature, {E}instein Metrics, and {S}eiberg-{W}itten Theory.
\newblock {\em Math. Res. Lett.} {\bf 1998}, {\em 5},~423--438.

\bibitem[Asselmeyer-Maluga and Kr{\'o}l(2014)]{AsselmeyerKrol2014}
Asselmeyer-Maluga, T.; Kr{\'o}l, J.
\newblock Inflation and Topological Phase Transition Driven by Exotic
  Smoothness.
\newblock {\em Advances in HEP} {\bf 2014}, {\em Article ID 867460},~14 pages.


\bibitem[Asselmeyer-Maluga and Krol(2018)]{AsselmeyerKrol2018c}
Asselmeyer-Maluga, T.; Krol, J.
\newblock A Topological Model for Inflation.
\newblock arXiv:1812.08158, subm. to Phys. Rev. D.

\bibitem[Asselmeyer-Maluga(2019)]{AsselmeyerMaluga2018d}
Asselmeyer-Maluga, T.
\newblock Hyperbolic Groups, 4-Manifolds and Quantum Gravity.
\newblock {\em Journal of Physics: Conference Series} {\bf 2019}, {\em {\bf
  1194}},~012009.


\bibitem[Mostow(1968)]{Mos:68}
Mostow, G.
\newblock Quasi-conformal mappings in $n$-space and the rigidity of hyperbolic
  space forms.
\newblock {\em Publ. Math. IHES} {\bf 1968}, {\em {\bf 34}},~53--104.

\bibitem[Prasad(1973)]{Prasad:1973}
Prasad, G.
\newblock Strong Rigidity of {\bf Q}-Rank 1 Lattices.
\newblock {\em Inv. Math.} {\bf 1973}, {\em {\bf 21}},~255--286.

\bibitem[Asselmeyer-Maluga and Kr{\'o}l(2014)]{AsselmeyerKrol2009}
Asselmeyer-Maluga, T.; Kr{\'o}l, J.
\newblock Abelian Gerbes, Generalized Geometries and Foliations of Small Exotic
  {$R^4$}.
\newblock arXiv: 0904.1276v5, subm. to Rev. Math. Phys.

\bibitem[Hurder and Katok(1984)]{HurKat:84}
Hurder, S.; Katok, A.
\newblock Secondary Classes and Trasnverse Measure Theory of a Foliation.
\newblock {\em BAMS} {\bf 1984}, {\em 11},~347 -- 349.
\newblock announced results only.

\bibitem[Serra and Dominguez~Romero(2011)]{DarkMatterEquation2011}
Serra, A.L.; Dominguez~Romero, M.J.L.
\newblock Measuring the Dark Matter Equation of State.
\newblock {\em Mon. Not. R. Astron. Soc.} {\bf 2011}, {\em {\bf 415}},~L74.


\bibitem[Sartoris(2014)]{DarkMatterEquation2014}
Sartoris, B.E.
\newblock {CLASH}-{VLT}: Constraints on the Dark Matter Equation of State from
  Accurate Measurements of Galaxy Cluster Mass Profiles.
\newblock {\em The Astrophysical Journal Letters} {\bf 2014}, {\em {\bf
  783}},~L11.


\bibitem[Kopp \em{et~al.}(2018)Kopp, Skordis, Thomas, and
  Ili{\'c}]{DarkMatterEquation2018}
Kopp, M.; Skordis, C.; Thomas, D.; Ili{\'c}, S.
\newblock Dark Matter Equation of State through Cosmic History.
\newblock {\em Phys. Rev. Lett. 120,} {\bf 2018}, {\em {\bf 120}},~221102.
\newblock arXiv:1802.09541.

\end{thebibliography}

\end{document}